\documentclass[11pt,fleqn]{article}
\usepackage{graphicx,cite,amsmath}

\newcommand{\eqn}[1]{(\ref{#1})}

\usepackage{subfigure}

\newcommand{\IM}{{\rm Im}}

\def\R1{\varepsilon_1}
\def\E8{\varepsilon_8}

\def\epe{\varepsilon'/\varepsilon}
\def\as{\alpha_s}

\newcommand{\gev}{\, {\rm GeV}}
\newcommand{\mev}{\, {\rm MeV}}
\newcommand{\bsi}{B_6^{(1/2)}}
\newcommand{\bei}{B_8^{(3/2)}}
\newcommand{\Lms}{\Lambda_{\overline{\rm MS}}}

\newcommand{\bea}{\begin{eqnarray}}
\newcommand{\eea}{\end{eqnarray}}
\newcommand{\bd}{\begin{displaymath}}
\newcommand{\ed}{\end{displaymath}}

\newcommand{\be}{\begin{equation}}
\newcommand{\ee}{\end{equation}}
\newcommand{\bi}{\begin{itemize}}
\newcommand{\ei}{\end{itemize}}
\newcommand{\ord}{{\cal O}}

\newcommand{\imlt}{\IM\lambda_t}

\begin{document}

\thispagestyle{empty}
\phantom{xxx}
\vspace{-24mm}
\begin{flushright}
 TUM-HEP-511/03 \\
 HD-THEP-03-26 \\[1cm]
\end{flushright}

\begin{center}

{\LARGE\bf \boldmath{$\epe$} at the NLO: 10 Years Later\\ }

\vskip1truecm
\centerline{\large\bf Andrzej J. Buras${}^a$ and
 Matthias Jamin${}^{b,*}$} 
\bigskip
{\sl ${}^a$ Physik Department, Technische Universit{\"a}t M{\"u}nchen}\\
{\sl D-85748 Garching, Germany}\\[2ex]
{\sl ${}^{b}$ Institut f\"ur Theoretische Physik, Universit\"at Heidelberg,}\\
{\sl Philosophenweg 16, D-69120 Heidelberg, Germany}\\[1cm]
\end{center}

\centerline{\bf Abstract}
During the last four years several parameters relevant for the analysis of 
the CP-violating ratio $\epe$ improved and/or changed significantly. In
particular, the experimental value of $\epe$ and the strange quark mass
decreased, the uncertainty in the CKM factor has been reduced, and for a
value of the hadronic matrix element of the dominant electroweak penguin
operator $Q_8$, some consensus has been reached among several theory groups.
In view of this situation, ten years after the first analyses of $\epe$ at
the next-to-leading order, we reconsider the analysis of $\epe$ within the 
SM and  investigate what can be said about the hadronic $Q_6$ matrix element 
of the dominant QCD penguin operator on the basis of the present experimental
value for $\epe$ and todays values of all other parameters.

Employing a conservative range for the reduced electroweak penguin matrix
element $R_8=1.0 \pm 0.2$ from lattice QCD, and present values for all other
input parameters, on the basis of the current world average for $\epe$, we
obtain the reduced hadronic matrix element of the dominant QCD penguin operator
$R_6=1.23\pm 0.16$ implying $\langle Q_6\rangle_0^{{\rm NDR}}(m_c) \approx 
-\,0.8\,\langle Q_8\rangle_2^{{\rm NDR}}(m_c)$. We compare these results with
those obtained in large--$N_c$ approaches in which generally $R_6\approx R_8$
and $\langle Q_6\rangle_0^{{\rm NDR}}(m_c)$ is chirally suppressed relatively
to $\langle Q_8\rangle_2^{{\rm NDR}}(m_c)$. We present the correlation between 
$R_6$ and $R_8$ that is implied by the data on $\epe$ provided new physics 
contributions to $\epe$ can be neglected.

\vfill


\vspace{4mm}
{\small ${}^{*}$ Heisenberg fellow.}

%%% end title page %%%%%%%%%%%%%

\newpage

\section{Introduction}
\setcounter{equation}{0}
The ratio $\epe$ that parametrises the size of direct CP violation with
respect to the indirect CP violation in $K_L\to \pi\pi$ decays has been the 
subject of very intensive experimental and theoretical studies in the last
three decades. After tremendous efforts, on the experimental side the world
average based on the recent results from NA48 \cite{NA4802,NA4801} and KTeV
\cite{KTeV03,KTeV99}, and previous results from NA31 \cite{NA3188,NA31} and
E731 \cite{E731}, reads
\begin{equation}
  \label{eps}
  \epe=(16.6\pm 1.6) \cdot 10^{-4} \qquad\qquad (2003)~.
\end{equation}
A recent discussion of the experimental issues involved in these analyses
can be found in ref.~\cite{Hol02}.

On the other hand, the theoretical estimates of this ratio are subject to
very large hadronic uncertainties. While several analyses of recent years
within the Standard Model (SM) find results that are compatible with \eqn{eps}
\cite{HPdR03,PPS01,BP01,PP01,BP00,BEF01,PP00,HKPS00,EPE99,HKPSB98}, it is
fair to say that even ten years after the first analyses of $\epe$ at the
next-to-leading order (NLO) \cite{BJL93,CFMR93}, the chapter on the theoretical
calculations of $\epe$ is certainly still open. A full historical account of
the theoretical efforts before 1998 can for example be found in the reviews
\cite{BFE98,BUR98}.
 
It should be emphasised that all existing analyses of $\epe$ use the NLO
Wilson coefficients calculated by the Munich and Rome groups in 1993
\cite{BJL93,CFMR93,BJLW92,BJLW93a,BJLW93b,CFMR94}, but the hadronic matrix
elements, the main theoretical uncertainty in $\epe$, vary from paper to paper.
Nevertheless, apart from the hadronic matrix element of the dominant QCD
penguin operator $Q_6$, in the last years progress has been made with the
determination of all other relevant parameters, which enter the theoretical
prediction of $\epe$. In view of this situation, we think it is legitimate
to reconsider the analysis of $\epe$ within the SM and to investigate what 
can be said about the hadronic $Q_6$ matrix element on the basis of the
present experimental value of $\epe$ and todays values of all other parameters.
In doing this we assume that any potential new physics contributions to $\epe$
can be neglected. We will return to this point at the end of our paper.

The phenomenological approach towards the values of the hadronic matrix 
elements presented in this work is a continuation of our approach in
\cite{BJL93}, where we extracted the matrix elements of current-current
operators $Q_i~(i=1-4)$ from the CP conserving data. In that case the
assumption about the absence of significant new physics contributions to
the relevant amplitudes was not necessary. On the contrary, the present
analysis has this particular limitation. Nevertheless, as we shall see below
some insight into the pattern of the size of the hadronic matrix elements of
penguin operators can be gained.

In order to set the scene for our investigation let us recall that after NA31
in 1993 \cite{NA31} and KTeV in 1999 \cite{KTeV99} presented surprisingly high
values for $\epe$, the 1999 world average including the E731 value \cite{E731}
was
\begin{equation}
  \label{eps99}
  \epe=(21.8\pm 3.0) \cdot 10^{-4} \qquad\qquad (1999)~,
\end{equation}
being roughly $30\%$ higher than the 2003 value. Within our 1999 analysis
of $\epe$ \cite{EPE99} that was performed using matrix elements of penguin
operators $Q_6$ and $Q_8$ in the ballpark of the values obtained in the
large--$N_c$ approach of \cite{BBG86,BBG87a,BBG87b}, it was essentially
impossible to reproduce the value in \eqn{eps99}. We found $\epe$ typically
a factor of two to three below the 1999 experimental value.

As we shall see below, the confrontation of the SM with the 2003 experimental 
value for $\epe$ as given in \eqn{eps}, does not resemble the 1999
situation presented by us in \cite{EPE99}. The following important facts are
responsible for this change:
\begin{itemize}
\item
The experimental value of $\epe$ decreased approximately by a factor of $1.3$.
\item
The value of $m_s$, relevant for $\epe$ within the approach of
\cite{BBG86,BBG87a,BBG87b}, decreased by roughly $15\%$, enhancing the
theoretical value of $\epe$ by roughly a factor of $1.3$.
\item
The isospin breaking parameter $\Omega_{{\rm IB}}$ decreased from 
roughly $0.25$ to $0.06$, enhancing $\epe$ by another factor of $1.4$.
\item
The  value for the hadronic matrix element of the dominant electroweak
penguin operator $\langle Q_8\rangle$ 
became
more robust due to general consistency of results from lattice QCD and other
analytic non-perturbative approaches to be discussed below.
\item
Finally, the value of the CKM factor $\IM\lambda_t$ became more precise,
although it did not change significantly compared to our 1999 value.
\end{itemize}

As a result of all these changes, our SM value for $\epe$ presented below
turns out to be fully consistent with the experimental value given in
\eqn{eps} for values of the hadronic matrix elements $\langle Q_6\rangle$
and $\langle Q_8\rangle$ which do not differ by more than $20\%$ from the
ones obtained within the large--$N_c$ approach of \cite{BBG86,BBG87a,BBG87b}.
On the other hand, the chiral suppression of $\langle Q_6\rangle_0$ with
respect to $\langle Q_8\rangle_2$ that is characteristic for the latter
approach, is only partially supported by the data.

In the next section, we recollect the basic equations which are required for
the analysis of $\epe$ within the SM. In section~3, we present the details
and numerical results of this work, in section~4 we compare our results with
those obtained in various  approaches, and in section~5 we summarise and
draw some conclusions.

\section{Basic Formulae}
\setcounter{equation}{0}
The central formula for $\epe$ of refs. \cite{EPE99,BRMSSM,BJL93} takes
the following form:
\be \frac{\varepsilon'}{\varepsilon}= \IM\lambda_t
\cdot F_{\varepsilon'}(x_t)
\label{epeth}
\ee
where
\be
\lambda_t=V^*_{ts}V_{td} ~, 
\qquad x_t = {\frac{m_t^2}{M_W^2}} 
\ee
and
\be
F_{\varepsilon'}(x_t) =P_0 + P_X \, X_0(x_t) + 
P_Y \, Y_0(x_t) + P_Z \, Z_0(x_t)+ P_E \, E_0(x_t)~.
\label{FE}
\ee
The functions $X_0$, $Y_0$, $Z_0$ and $E_0$ are given by
($m_t=165\gev$)
\begin{equation}\label{X0}
X_0(x_t)={\frac{x_t}{8}}\left[{\frac{x_t+2}{x_t-1}} 
+ {\frac{3 x_t-6}{(x_t -1)^2}}\; \ln x_t\right]= 1.505 ~,
\end{equation}
\begin{equation}\label{Y0}
Y_0(x_t)={\frac{x_t}{8}}\left[{\frac{x_t -4}{x_t-1}} 
+ {\frac{3 x_t}{(x_t -1)^2}} \ln x_t\right]= 0.962 ~,
\end{equation}
\bea\label{Z0}
Z_0(x_t)\!\!\!\!&=&\!\!\!\!-\,{\frac{1}{9}}\ln x_t + 
{\frac{18x_t^4-163x_t^3 + 259x_t^2-108x_t}{144 (x_t-1)^3}}+
\nonumber\\ 
&&\!\!\!\!+\,{\frac{32x_t^4-38x_t^3-15x_t^2+18x_t}{72(x_t-1)^4}}\ln x_t=0.664
\eea
\begin{equation}\label{E0}
E_0(x_t)=-\,{\frac{2}{3}}\ln x_t+{\frac{x_t^2(15-16x_t+4x_t^2)}{6(1-x_t)^4}}
\ln x_t+{\frac{x_t(18-11x_t-x_t^2)}{12(1-x_t)^3}}=0.271 ~.
\end{equation}

\begin{table}[thb]
\begin{center}
\begin{tabular}{|c||c|c|c||c|c|c||c|c|c|}
\hline
& \multicolumn{3}{c||}{$\Lms^{(4)}=310\mev$} &
  \multicolumn{3}{c||}{$\Lms^{(4)}=340\mev$} &
  \multicolumn{3}{c| }{$\Lms^{(4)}=370\mev$} \\
\hline
$i$ & $r_i^{(0)}$ & $r_i^{(6)}$ & $r_i^{(8)}$ &
      $r_i^{(0)}$ & $r_i^{(6)}$ & $r_i^{(8)}$ &
      $r_i^{(0)}$ & $r_i^{(6)}$ & $r_i^{(8)}$ \\
\hline
0 &
   --3.574 &  16.552 &   1.805 &
   --3.602 &  17.887 &   1.677 &
   --3.629 &  19.346 &   1.538 \\
$X_0$ &
     0.574 &   0.030 &       0 &
     0.564 &   0.033 &       0 &
     0.554 &   0.036 &       0 \\
$Y_0$ &
     0.403 &   0.119 &       0 &
     0.392 &   0.127 &       0 &
     0.382 &   0.134 &       0 \\
$Z_0$ &
     0.714 &  --0.023 &  --12.510 &
     0.766 &  --0.024 &  --13.158 &
     0.822 &  --0.026 &  --13.855 \\
$E_0$ &
     0.213 &  --1.909 &   0.550 &
     0.202 &  --2.017 &   0.589 &
     0.190 &  --2.131 &   0.631 \\
\hline
\end{tabular}
\end{center}
\caption[]{The coefficients $r_i^{(0)}$, $r_i^{(6)}$ and $r_i^{(8)}$ of
formula \eqn{eq:pbePi} for various $\Lms^{(4)}$ in the NDR scheme.
\label{tab:pbendr}}
\end{table}

The coefficients $P_i$ are given in terms of the non-perturbative parameters
$R_6$ and $R_8$ as follows:
\begin{equation}
P_i = r_i^{(0)} + 
r_i^{(6)} R_6 + r_i^{(8)} R_8 \,,
\label{eq:pbePi}
\end{equation}
where the coefficients $r_i^{(0)}$, $r_i^{(6)}$ and $r_i^{(8)}$ comprise
information on the Wilson-coefficient functions of the $\Delta S=1$ weak
effective Hamiltonian at the next-to-leading order
\cite{BJLW92,BJLW93a,BJLW93b,CFMR94}, and their numerical values for
different values of $\Lms^{(4)}$ at $\mu=m_c$ in the NDR renormalisation
scheme are displayed in table~\ref{tab:pbendr}. This table updates the
corresponding ones presented in \cite{EPE99,BRMSSM}.

In terms of the hadronic $B$-parameters $\bsi$ and $\bei$, as well as the
strange quark mass, the non-perturbative parameters $R_6$ and $R_8$ are
defined as
\be\label{RS}
R_6\equiv \bsi\left[ \frac{121\mev}{m_s(m_c)+m_d(m_c)} \right]^2,
\qquad
R_8\equiv \bei\left[ \frac{121\mev}{m_s(m_c)+m_d(m_c)} \right]^2.
\ee
In the large--$N_c$ approach of \cite{BBG86,BBG87a,BBG87b}, the matrix
elements of the dominant QCD-penguin ($Q_6$) and the dominant electroweak
penguin ($Q_8$) operator, are then given by \cite{BBH90,BJL93}
\begin{eqnarray}
\label{Q6fac}
\langle Q_6 \rangle_0 \!\!\!&=&\!\!\! -\,4\,\sqrt{\frac{3}{2}}\,(F_K-F_\pi)
\biggl(\!\frac{m_K^2}{121\mev}\!\biggr)^{\!\!2} R_6 \,=\,
-\,0.597\cdot R_6\,\gev^3 \,, \\
\label{Q8fac}
\langle Q_8 \rangle_2 \!\!\!&=&\!\!\! \sqrt{3}\,F_\pi
\biggl(\!\frac{m_K^2}{121\mev}\!\biggr)^{\!\!2} R_8 \,=\,
0.948\cdot R_8\,\gev^3 \,,
\end{eqnarray}
where for $\langle Q_8 \rangle_2$, we have neglected a small 1\% correction
which is independent of $m_s$, and the numerical values have been obtained by
employing values for $F_\pi$, $F_K$ and $m_K$ as given in the current PDG
\cite{PDG02}. Let us remark that the overall normalisation of our matrix
elements in \eqn{Q6fac} and \eqn{Q8fac} differs by a factor $\sqrt{3/2}$ from
another convention used in the literature. This factor, however, compensates
with a corresponding factor in our definition of the isospin amplitudes to
provide the same physical results.

In the strict large--$N_c$ limit, $B_6^{(1/2)}=B_8^{(3/2)}=1$, the
$\mu$--dependence of the matrix elements $\langle Q_6 \rangle_0$ and
$\langle Q_8 \rangle_2$ is fully governed by the $\mu$--dependence of the quark
masses $m_{s,d}(\mu)$, and, as the mixing with other operators can be neglected
in this limit, the evolution of the Wilson coefficients $C_6(\mu)$ and
$C_8(\mu)$ precisely cancels the $\mu$--dependence of $\langle Q_6 \rangle_0$
and $\langle Q_8 \rangle_2$ such that $\epe$ is $\mu$--independent as it should
be. As our numerical analysis of ref.~\cite{BJL93} demonstrated, $B_6^{(1/2)}$
and $B_8^{(3/2)}$ remain independent of $\mu$ to a very good accuracy also
outside the strict large-$N_c$ limit, provided these parameters do not deviate
considerably from unity.

The formulae \eqn{RS}, together with the strict large--$N_c$ limit, also
imply $R_6=R_8$, so that in this limit there is a one-to-one correspondence
between $\epe$ and $R_6=R_8$ for fixed values of the remaining parameters.
Consequently, in this case, only for certain values of $m_s(m_c)$ is one able
to obtain the experimental value for $\epe$. This has been emphasised in
particular in \cite{KNS99}, and we shall return to this point in the next
section. Thus in this approach a concrete prediction for $\epe$ can be made
provided the value $m_s(m_c)$ has been determined by some other method.
Moreover, once $m_s(m_c)$ is known, also $R_6$, $R_8$, $\langle Q_6 \rangle_0$
and $\langle Q_8 \rangle_2$ are known, but they always satisfy the relations:
\be\label{REL}
\frac{R_6}{R_8}=1~, \qquad
\frac{\langle Q_6 \rangle_0}{\langle Q_8 \rangle_2}=-\,0.63~.
\ee
We shall compare the relations \eqn{REL} with our results for
$\langle Q_6 \rangle_0$ and $\langle Q_8 \rangle_2$ obtained by various
methods below.

\section{ Numerical Analysis}
\setcounter{equation}{0}
\subsection{Preliminaries}

Before we come to a detailed presentation of our numerical analysis, let us
first briefly discuss the origin of the values for the dominant input
parameters that we have used in our analysis.

The dependence of $\epe$ on the quark mixing or CKM matrix elements appears
through the parameter $\IM\lambda_t$. Fits determining the elements of the
CKM matrix have been the subject of extensive efforts in the last years. The
present status has been summarised very recently in the review \cite{CKM03}.
Employing the best fit incorporating all known constraints within the SM,
the resulting value for $\IM\lambda_t$ is then found to be \cite{Sto03}:
\begin{equation}
\IM\lambda_t = (1.31 \pm 0.10)\cdot 10^{-4}~.
\end{equation}

The next parameter which deserves an explicit mentioning since $\epe$ is
rather sensitive to its precise value, is the quantity $\Omega_{{\rm IB}}$,
parametrising isospin-breaking corrections which result from strong isospin
violation ($m_u\neq m_d$) and electromagnetic corrections. For this parameter,
we employ the result of the very recent analysis \cite{CPEN03}
(see also refs. \cite{EMNP00,MW01}): $\Omega_{{\rm IB}} \,=\, 0.06 \pm 0.08$.
The shift from $\Omega_{{\rm IB}} \,=\, 0.25$, that was used in our previous
analysis of $\epe$ \cite{EPE99}, to the new result, has the largest impact
for the theoretical value of $\epe$.

An important fundamental QCD input parameter is the value of the strong
coupling constant $\as$. As pointed out in refs.~\cite{BL93,EPE99}, $\epe$
is roughly proportional to the QCD parameter $\Lms$ in a four-flavour theory,
when this parameter is $\ord(350\mev)$. For the strong coupling, we have used
the recent average $\as(M_Z)=0.119\pm 0.002$, corresponding to
$\Lms^{(4)}=340\pm 30\,\mev$.

Another parameter is the top quark mass that enters the Inami-Lim functions
given above and the fit of $\IM\lambda_t$. However, $m_t$ is already well
known and its precise value is less important. On the other hand, the value
of the strange quark mass $m_s$ plays an important role in the large-$N_c$
approach discussed in the previous section. For the quark masses, we shall
then use the following values:
\begin{equation}\label{msmt}
m_t(m_t) = 165 \pm  5 \,\gev~, \qquad
m_s(m_c) = 115 \pm 20 \,\mev~.
\end{equation}
The top mass evaluated at the scale $m_t$ results from converting the
corresponding pole mass value \cite{PDG02}, and the strange quark mass is
an average over recent determinations of this quantity
\cite{GJPPS03,CDGHPP01,JOP02,MK02,Wit02}. Due to our method of extracting
the matrix elements of current-current operators $Q_i~(i=1-4)$ from the CP
conserving data \cite{BJL93}, the strange mass is required at the scale
$m_c=1.3\,\gev$ and our central value would correspond to
$m_s(2\,\gev)=100\,\mev$.

As stressed by several authors in the literature, the strong $m_s$ dependence
of $\langle Q_6 \rangle_0$ and $\langle Q_8 \rangle_2$ given in the
large--$N_c$ approach is somewhat artificial. In fact, the lattice studies
\cite{CM01} and the discussion in refs.~\cite{BP00,Don01,HPdR03} indicate that
these matrix elements are only weakly dependent on the actual value of $m_s$
since they are related to the quark condensate via the GMOR relation
\cite{GMOR68,Jam02}. Here we shall follow first a different approach. We will
directly deal with the parameters $R_6$ and $R_8$ of eq. \eqn{RS}, and only
at the end of our numerical analysis, we shall discuss the implications for
$B_6^{(1/2)}$ and $B_8^{(3/2)}$.

\subsection{Present knowledge of \boldmath{$R_8$}}
As can be inferred from the formulae presented in the last section, in order
to be able to calculate $\epe$, we still require some input for $R_6$ and
$R_8$ which parametrise the dominant contributions from QCD and electroweak
penguin operators respectively. Concerning the parameter $R_8$, in the last
years progress has been achieved both in the framework of lattice QCD
\cite{Blum01,Aoki01,Bec02} as well as with analytic methods to tackle QCD
non-perturbatively \cite{Nar00,KPdR01,BGP01,CDGM03,HPdR03}.

The current situation of the status of $\langle Q_8\rangle_2$ calculations has
been summarised nicely in the recent article \cite{CDGM03}. At present, the
most precise determination of $\langle Q_8\rangle_2$ comes from the lattice QCD
measurement \cite{Bec02}, corresponding to $R_8=1.0 \pm 0.1$. The lattice
studies of refs.~\cite{Blum01,Aoki01} find results in complete agreement,
however, with somewhat larger uncertainties. The analytic non-perturbative
approaches of refs.~\cite{Nar00,KPdR01,BGP01,CDGM03}, employing either the
large-$N_c$ expansion or dispersion relations, on the other hand, find results
that are generally larger than the lattice findings, although certainly
compatible within the uncertainties. However, these analyses have been
performed in the chiral limit, and as was demonstrated in \cite{CDGM03},
chiral corrections tend to decrease $R_8$. Therefore, in order to be more
conservative, and in view of the fact that there could be additional effects
of chiral logarithms when relating the matrix elements computed on the
lattice to the physically relevant ones, in what follows we shall use:
\be
\label{R8}
R_8=1.0 \pm 0.2~, \qquad
\langle Q_8\rangle_2^{{\rm NDR}}(m_c) = (0.95 \pm 0.19)\,\gev^3~.
\ee

\begin{table}[htb]
\renewcommand{\arraystretch}{1.2}
\begin{center}
\begin{tabular}{ccccccccc}
\hline
$R_8$ & $R_6$ & $\varepsilon'/\varepsilon$ & $\IM\lambda_t$ &
$\Lms$ & $\Omega_{{\rm IB}}$ & $m_t$ & $m_s$ & total \\
\hline
0.8 & 1.143 & ${}^{+0.070}_{-0.070}$ & ${}^{-0.051}_{+0.060}$
    & ${}^{-0.061}_{+0.065}$ & ${}^{+0.092}_{-0.078}$ & ${}^{+0.018}_{-0.017}$
    & ${}^{+0.007}_{-0.012}$ & ${}^{+0.147}_{-0.133}$ \\
1.0 & 1.222 & ${}^{+0.070}_{-0.070}$ & ${}^{-0.051}_{+0.060}$
    & ${}^{-0.061}_{+0.064}$ & ${}^{+0.099}_{-0.084}$ & ${}^{+0.024}_{-0.023}$
    & ${}^{+0.007}_{-0.012}$ & ${}^{+0.152}_{-0.138}$ \\
1.2 & 1.301 & ${}^{+0.070}_{-0.070}$ & ${}^{-0.051}_{+0.060}$
    & ${}^{-0.061}_{+0.064}$ & ${}^{+0.107}_{-0.090}$ & ${}^{+0.029}_{-0.028}$
    & ${}^{+0.007}_{-0.012}$ & ${}^{+0.158}_{-0.142}$ \\
2.2 & 1.695 & ${}^{+0.070}_{-0.070}$ & ${}^{-0.051}_{+0.060}$
    & ${}^{-0.059}_{+0.062}$ & ${}^{+0.143}_{-0.121}$ & ${}^{+0.058}_{-0.057}$
    & ${}^{+0.007}_{-0.012}$ & ${}^{+0.190}_{-0.170}$ \\
\hline
\end{tabular}
\end{center}
\caption{$R_6$ as a function of $R_8$ for four relevant values. In addition,
we give the variation of $R_6$ while varying all other input parameters. For
a more detailed explanation see the text.
\label{tab2}}
\end{table}
\subsection{Determining \boldmath{$R_6$}}
The situation is less clear concerning the QCD penguin contribution
parametrised by $R_6$. Therefore, we postpone a discussion of the present
status of $R_6$, and we shall take a different approach. Putting together
all the information presented thus far, let us investigate what can be said
about our expectations for $R_6$. A summary of this analysis is presented
in table~\ref{tab2}. The second column shows central values for $R_6$
corresponding to the four inputs for $R_8$ as given in the first column.
The remaining columns display the variation of $R_6$ while varying the
dominant input parameters, and the last column contains the total uncertainty
when all individual errors are added in quadrature.

\begin{figure}[htb]
\begin{center}
\includegraphics[angle=270, width=14cm]{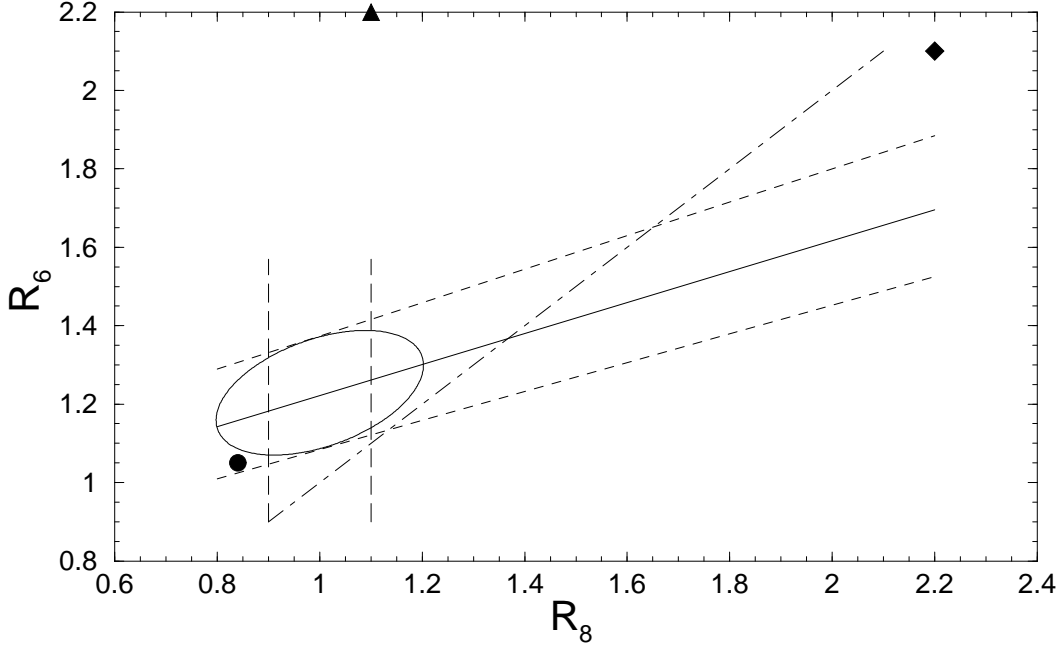}
\end{center}
\caption{$R_6$ as a function of $R_8$. For a detailed explanation see
the text.
\label{fig1}}
\end{figure}
A graphical representation of our results of table~\ref{tab2} in the spirit
of the ``$\epe$--path" of ref.~\cite{BG01} (see also refs.~\cite{Don01,HPdR03})
is given in figure~\ref{fig1}. The solid straight line corresponds to the
central values given in table~\ref{tab2}, whereas the short-dashed lines are
the uncertainties due to a variation of the input parameters. (Let us note
that all our errors should be considered as 1$\sigma$ deviations.) Next, the
vertical long-dashed lines indicate the lattice range for $R_8$ \cite{Bec02},
whereas the ellipse describes the correlation between $R_6$ and $R_8$ implied
by the data on $\epe$ when taking into account the more conservative constraint
on $R_8$ given in eq.~\eqn{R8}. From the allowed region within the ellipse, we
can infer our central result for $R_6$, which reads:
\begin{equation}
\label{R6}
R_6 = 1.23\pm 0.16~, \qquad
\langle Q_6\rangle_0^{{\rm NDR}}(m_c) = -\,(0.73 \pm 0.10)\,\gev^3~.
\end{equation}
Finally, the full triangle, circle and diamond in figure~1 represent the
central results of refs.~\cite{BP00,BGP01}, \cite{PPS01} as well as
\cite{HPdR03} respectively, which will be discussed in more detail below,
and the dashed-dotted line shows the strict large-$N_c$ relation $R_6=R_8$.

Comparing \eqn{R6} with \eqn{R8}, we observe that if the matrix element 
of $Q_8$ is indeed given by \eqn{R8} than the data on $\epe$ seem to require 
\begin{equation}
\label{Q6Q8}
\frac{R_6}{R_8}=1.23\pm0.29~, \qquad 
\langle Q_6\rangle_0^{{\rm NDR}}(m_c) = 
-\,(0.77\pm0.19)\,\langle Q_8\rangle_2^{{\rm NDR}}(m_c)~.
\end{equation}
Due to substantial uncertainties, these results are compatible with the
large--$N_c$ expectations in \eqn{REL} but the strong chiral suppression of
the matrix element $\langle Q_6\rangle_0^{{\rm NDR}}(m_c)$ with respect to 
$\langle Q_8\rangle_2^{{\rm NDR}}(m_c)$ found in the latter approach, although
visible in (\ref{Q6Q8}), is not fully supported by the data.
However, we would like to emphasise that this result is a consequence of
\eqn{R8}. We will return to this point in the next section, where we shall
compare the results in \eqn{R8}, \eqn{R6} and \eqn{Q6Q8} with other approaches.

\subsection{The dependence of \boldmath{$\epe$} on \boldmath{$R_6$} and
\boldmath{$R_8$}}
In addition, it is also of interest to show the values of $\epe$ for specific 
values of $R_6$, $R_8$ and $\Lms^{(4)}$. We do this in table~\ref{tab:eps}.
The results in table~\ref{tab:eps} should be self explanatory. Depending on
the values of the parameters $\Lms^{(4)}$, $R_6$ and $R_8$, one finds values
of $\epe$ that are within or outside the one standard deviation experimental
range given in \eqn{eps}. However, the remarkable feature of the values in
this table is that most of them are within two standard deviations from the
experimental value.

\begin{table}[htb]
\begin{center}
\begin{tabular}{|c||c|c||c|c||c|c|}
\hline
& \multicolumn{2}{c||}{$R_6 = 1.0$} &
  \multicolumn{2}{c||}{$R_6 = 1.2$} &
  \multicolumn{2}{c| }{$R_6 = 1.4$} \\
\hline
$\Lms^{(4)}~[\mev] $ & 
$R_8=0.8$ & $R_8=1.0$ &
$R_8=0.8$ & $R_8=1.0$ &
$R_8=0.8$ & $R_8=1.0$  \\
\hline
310 & 12.2 & 10.5 & 16.4 & 14.8 & 20.7 & 19.0 \\
340 & 13.3 & 11.5 & 17.9 & 16.1 & 22.5 & 20.7 \\
370 & 14.6 & 12.6 & 19.5 & 17.6 & 24.5 & 22.5 \\
\hline
\end{tabular}
\end{center}
\caption[]{The ratio $\epe$ in units of $10^{-4}$ for $m_t = 165 \gev$ 
and various values of $\Lms^{(4)}$, $R_6$, $R_8$. 
\label{tab:eps}}
\end{table}

\section{Comparison with other Analyses}
\setcounter{equation}{0}
Let us now  compare  our findings of \eqn{R8}, \eqn{R6} and \eqn{Q6Q8} with
other recent estimates of these quantities. 

In the approach of refs. \cite{PP00,PP01,PPS01}, the enhancement of the QCD
penguin matrix element $\langle Q_6\rangle_0$ above the factorisation result
was attributed to the strong final-state interactions in the final state pion
system. In \cite{PP00,PP01}, the final-state interactions were taken into
account through dispersion relation techniques which resulted in an Omn\'es
type exponential, giving an enhancement factor $1.55\pm 0.10$ which provides
full agreement with the experimental result for $\epe$.\footnote{Further
discussion of this approach can also be found in refs.~\cite{Buretal00,BCKO01}.}
No such enhancement has been found in the case of $\langle Q_8\rangle_2$, for
which a mild suppression factor $0.92\pm 0.03$ was obtained \cite{PPS01}.
Furthermore, the results of the Omn\'es approach were corroborated by
an explicit computation within chiral perturbation theory at one loop.

However, the strange mass used in ref. \cite{PPS01} corresponds to
$(m_s+m_d)(m_c)=135\,\mev$, thus being 10\% higher than the value advocated in
the present work. Translating the findings of \cite{PPS01} for the case of
Omn\'es resummed final-state interactions into $R_6$ and $R_8$, we obtain
\begin{equation}
\label{PP}
R_6 = 1.05 \pm 0.06~, \qquad 
R_8 = 0.84 \pm 0.05~,
\end{equation}
in reasonable agreement with our values, albeit pointing to a slightly
different scenario as to their composition into $B$-parameter and strange mass.
With the present uncertainty in the strange quark mass, which is not included
in the errors given in eq.~\eqn{PP}, we are, however, unable to decide which
is the correct scenario. The central result of \eqn{PP} is also indicated as
the full circle in figure \ref{fig1}.

In the very recent article \cite{HPdR03}, the matrix element of the QCD
penguin operator $Q_6$ was calculated within an approach which was able
to take into account the dominant ${\cal O}(n_f/N_c)$ corrections to the
large--$N_c$ result, where $n_f$ is the number of active quark flavours.
In addition, the authors demonstrated an explicit cancellation of the
renormalisation scale dependence between Wilson coefficients and matrix 
elements. Numerically, in our language, the results of \cite{HPdR03} and
\cite{KPdR01} would correspond to 
\begin{equation}
\label{EDU}
R_6 = 2.1 \pm 1.1~, \qquad 
R_8 = 2.2 \pm 0.4~.
\end{equation}
As seen in figure~\ref{fig1}, the central result, displayed as the full
diamond, is roughly consistent with the experimental value for $\epe$, but lies
far outside the ellipse found here. However, in the approach of \cite{HPdR03},
the dominant uncertainty for $R_6$ is due to the value of the quark condensate,
and, as already remarked above, both parameters are presently only calculated
in the chiral limit. Thus, in our opinion, at the moment it is premature to
draw any definite conclusions from the results of \cite{KPdR01} and
\cite{HPdR03}.

Similar results of a rather large value of the matrix element
$\langle Q_6\rangle_0$ were also found earlier in the framework of the $1/N_c$
expansion at NLO with chiral perturbation theory at the leading order
\cite{BP00,BP01}. With the value $m_s(2\,\gev)=105\,\mev$ advocated there,
the finding $\bsi=2.5\pm 0.4$ leads to $R_6=2.2\pm 0.4$, in perfect agreement
to the result of eq.~\eqn{EDU}. Together with the result $R_8=1.1\pm0.3$
\cite{BGP01}, in figure~1 the findings of \cite{BP00,BP01,BGP01} are displayed
as the full triangle. It should be clear that this set of parameters leads to
a value of $\epe$ much larger than the experimental average. Nevertheless, one
should note that like in ref. \cite{HPdR03}, also in \cite{BP00,BP01}, the
penguin matrix element $\langle Q_6\rangle_0$ was expressed in terms of the
quark condensate. It is, however, well known that the chiral corrections to
the GMOR relation which relates the parametrisations \eqn{Q6fac}, and the one
in terms of the quark condensate, in the strange quark case are large, of order
50\% \cite{Jam02}. This can easily lead to a factor of two difference for both
parametrisations, and one has to wait for a complete computation at the
next-to-leading order, before it is clear which one is closer to the true
result.

%Still, the interesting property of the result in \eqn{EDU} is its compatibility
%with the large--$N_c$ limit relation in \eqn{REL}. With $B_6^{(1/2)} =
%B_8^{(3/2)}=1$, one would need $m_s(m_c)\approx 105\mev$ or equivalently
%$m_s(2\gev)\approx 90\,\mev$, which is somewhat smaller than \eqn{msmt},
%but lower values for $m_s$ have also been found in lattice simulations with
%dynamical fermions \cite{Wit02,HCLMT02,Gup03}.

To conclude this section, we further elaborate on the strict large--$N_c$ limit
and perform the following exercise. We set
\be
\bsi=\bei=1.0 ~\mbox{\cite{BBG86,BG87}}
\quad \mbox{and} \quad
\hat B_K=0.75~\mbox{\cite{GTW81,BG86}},
\ee
In that case one finds
\cite{Sto03}
\be
\imlt=(1.34\pm 0.06)\cdot 10^{-4}~.
\ee
Setting $\imlt$ to this central value, in table~\ref{tab:LN} we show the
results for $\epe$ as a function of $m_s(m_c)$ and $\Lms^{(4)}$. 
\begin{table}[htb]
\begin{center}
\begin{tabular}{|c||c|c|c|}
\hline
$\Lms^{(4)}~[\mev] $ & $m_s(m_c)=115\mev$ & $m_s(m_c)=105\mev$ &
                       $m_s(m_c)= 95\mev$ \\
\hline
310 & 10.8 & 13.4 & 16.8  \\
340 & 11.8 & 14.6 & 18.2  \\
370 & 12.9 & 15.9 & 19.9 \\
\hline
\end{tabular}
\end{center}
\caption[]{The ratio $\epe$ in units of $10^{-4}$ for the strict large--$N_c$
results $\bsi=\bei=1.0$, $\hat B_K=0.75$ and various values of $\Lms^{(4)}$
and $m_s(m_c)$. 
\label{tab:LN}}
\end{table}

We observe that for $m_s(m_c)\approx 105\mev$ or equivalently
$m_s(2\gev)\approx 90\mev$ and $\Lms^{(4)}\approx 340\mev$, the results in table
\ref{tab:LN} are fully consistent with the data on $\epe$. The required value
of $m_s$ is smaller, but consistent with \eqn{msmt} and in the ballpark of
values found in the most recent lattice simulations with dynamical fermions
\cite{Wit02,HCLMT02,Gup03}. As can be seen from the dashed-dotted line in
figure~\ref{fig1}, the required value for the large--$N_c$
condition $R_6=R_8$ is 
\begin{equation}
\label{BJ}
R_6 =R_8= 1.36\pm 0.3~,
\end{equation}
which is close to the ellipse found here but outside of it.

\section{Conclusions}
\setcounter{equation}{0}
In this letter, we have reanalysed the ratio $\epe$ in view of the improved
experimental data and an improved information on the relevant theoretical input
parameters. After the uncertainties on $\imlt$ have been reduced significantly
during the last four years and some consensus has been reached on the matrix
element of the dominant electroweak penguin operator $Q_8$, the main
uncertainty in the estimate of $\epe$ in the SM is the matrix element of the
leading QCD penguin operator $Q_6$. Assuming that $\epe$ is fully dominated
by the SM contributions, we have determined the ratio $R_6$ 
and the matrix element $\langle Q_6\rangle_0^{{\rm NDR}}(m_c)$
from the data on $\epe$ with the results given in (\ref{R6}) and
(\ref{Q6Q8}).

We have also shown that with $\hat B_K=0.75$ and $m_s(m_c)\approx 105\mev$, 
the large--$N_c$ values of $\langle Q_6\rangle_0$ and $\langle Q_8\rangle_2$ 
with their Wilson coefficients calculated within the SM are consistent with
the experimental data on $\epe$. This should be contrasted with the situation
of 1999, when the large--$N_c$ estimates of $\epe$ were by a factor of two to
three lower than the data. The reasons for this change have already been
listed in section~1. 

Finally, we have shown in figure~\ref{fig1} the correlation between $R_6$
and $R_8$ as implied by the experimental data on $\epe$.

There are three messages from our paper:
\begin{itemize}
\item
If indeed $R_8=1.0\pm 0.2$ as indicated by several recent estimates, then
the data on $\epe$ imply
\begin{equation}
\label{Q6Q8a}
R_6 = 1.23 \pm 0.16~, \quad
\frac{R_6}{R_8}= 1.23\pm0.29~, \quad 
\langle Q_6\rangle_0^{{\rm NDR}}(m_c)= 
-\,(0.77\pm0.19)\,\langle Q_8\rangle_2^{{\rm NDR}}(m_c)~.
\end{equation}
This is in accordance with the results in \cite{PP00,PP01,PPS01}, but as seen
in figure~\ref{fig1} it   differs 
from the large--$N_c$ approaches in \cite{BBG86,BBG87a,BBG87b}, 
\cite{BP00,BP01,BGP01} and
\cite{HPdR03}  with the first one closest to our findings. 
%in which $R_6\approx R_8$ and
%$\langle Q_6\rangle_0^{{\rm NDR}}(m_c)$ is chirally suppressed relatively to 
%$\langle Q_8\rangle_2^{{\rm NDR}}(m_c)$.
\item
The large--$N_c$ approach of \cite{BBG86,BBG87a,BBG87b} can only be made
consistent with data provided 
\begin{equation}
R_6=R_8=1.36\pm 0.3
\end{equation}
and $\langle Q_8\rangle_2^{{\rm NDR}}(m_c)$ is larger than obtained by most
recent approaches discussed above. This would require
$m_s(m_c)\approx 105\mev$ which is still compatible with \eqn{msmt}. Moreover,
lower values of $m_s(m_c)$ are indicated by the recent lattice simulations
with dynamical fermions \cite{Wit02,HCLMT02,Gup03}, but the situation
certainly requires further study.
\item
Higher values of 
$\langle Q_6\rangle_0^{{\rm NDR}}(m_c)$ and
$\langle Q_8\rangle_2^{{\rm NDR}}(m_c)$ are obtained in the approach of 
\cite{HPdR03} but as the calculations have been done in the chiral 
limit, it would be premature to draw definite conclusions at present.
 From the point of view of the approach in \cite{BBG86,BBG87a,BBG87b} 
the values $R_6\approx R_8\approx 2$ would correspond to rather 
small values of $m_s(m_c)$ so that a better interpretation would be a
significant departure of the parameters $B_6^{(1/2)}$ and $B_8^{(3/2)}$ from
unity. As the analysis \cite{HPdR03} goes beyond the strict large--$N_c$ limit,
this is certainly conceivable.
\end{itemize}

As shown  in figure~\ref{fig1}, all these three scenarios are consistent 
with the data. Which of these pictures of $\epe$ is correct, can only be 
answered by calculating $\langle Q_6\rangle_0$ by means of non-perturbative 
methods that go beyond the large--$N_c$ limit and are reliable. Moreover, the
calculation of the strange quark mass $m_s$ and of $\langle Q_8\rangle_2$
should be improved. Such calculations are independent of the assumption about
the role of new physics in $\epe$ that we have made in order to extract
$\langle Q_6\rangle_0$ from the data.

If the values for these matrix elements and $R_{6,8}$ will be found one day 
to lie significantly outside the allowed region in figure~\ref{fig1}, new
physics contributions to $\epe$ will be required in order to fit the
experimental data.

%******************************************************************************
\section*{Acknowledgements}
%******************************************************************************

M.J. would like to thank D. Becirevic, S. Peris and J. Prades for discussions.
This research was partially supported by the German `Bundesministerium f\"ur
Bildung und Forschung' under contract 05HT1WOA3 and by the `Deutsche
Forschungsgemeinschaft' (DFG) under contract Bu.706/1-2. M.J. would also
like to thank the Deutsche Forschungsgemeinschaft for support.

\newpage
%\bibliography{bj03}

\end{document}